# An intelligent sociotechnical systems (iSTS) framework: Enabling a hierarchical human-centered AI (hHCAI) approach

Wei Xu, *Senior Member, IEEE;* Zaifeng Gao

***Abstract*—While artificial intelligence (AI) offers significant benefits, it also has negatively impacted humans and society. A human-centered AI (HCAI) approach has been proposed to address these issues. However, current HCAI practices have shown limited contributions due to a lack of sociotechnical thinking. To overcome these challenges, we conducted a literature review and comparative analysis of sociotechnical characteristics with respect to AI. Then, we propose updated sociotechnical systems (STS) design principles. Based on these findings, this paper introduces an intelligent sociotechnical systems (iSTS) framework to extend traditional STS theory and meet the demands with respect to AI. The iSTS framework emphasizes human-centered joint optimization across individual, organizational, ecosystem, and societal levels. The paper further integrates iSTS with current HCAI practices, proposing a hierarchical HCAI (hHCAI) approach. This hHCAI approach offers a structured approach to address challenges in HCAI practices from a broader sociotechnical perspective. Finally, we provide recommendations for future iSTS and hHCAI work.**

## I. INTRODUCTION

ARTIFICIAL intelligence (AI)-based intelligent systems have brought benefits to humans and society. While we benefit from AI, research indicates that its limitations can negatively impact both humans and society [1], [2], [3], [4], [5], [6], [7], such as issues related to vulnerability, trust, bias, fairness, explainability, causal models, and ethical AI. Yampolskiy et al. [8] reported that improper development and use of AI technology have led to many accidents that impact human fairness, justice, and safety. The AI Accident Database (AIID) [9] has collected over a thousand AI-related accidents, including autonomous vehicles hitting and killing pedestrians, trading algorithm errors leading to market "flash crashes," facial recognition system errors leading to the arrest of innocent people, and so on. The AIAAIC database [10] tracking AI abuse incidents shows that the number of incidents has increased 26-fold since 2012. The adverse effects of AI technology may occur when using various types of intelligent systems. For example, some intelligent systems trained with distorted data may cause the system to produce bias, easily amplify prejudice and unfairness, and so on, and the "world view" they follow may put certain user groups at a disadvantage and affect social fairness, causing some AI projects to fail [8], [11], [12], [13]. When more and more organizations adopt intelligent decision-making systems, decisions generated by AI systems with biased "world views" will directly affect people's daily work and lives.

It is evident that there is a double-edged sword effect for AI technology, meaning that rational use of AI will benefit humans, and unreasonable use and algorithmic fallout will harm humans and society [14], [15]. We can see this demonstrated in the fields of nuclear energy and biotechnology. The rational use of nuclear power and biochemical technology can benefit humans, but misuse may cause disastrous consequences. However, nuclear power and biotechnology are relatively challenging to master and are highly centralized technologies. The development and use of AI technology is a decentralized global phenomenon, and the entry threshold is relatively low, which makes it more challenging and critical to control [16]. Therefore, we must develop effective strategies and approaches to address its negative impact on humans and society.

To address the challenges brought about by AI, a human-centered AI (HCAI) approach has been proposed to address the ignorance of humans and society as a priority in the current technology-driven approach to developing and deploying AI systems [1], [3], [17], [18], [19], [20], [21], [22], [23], [24]. For example, Shneiderman [1] and Xu [2] specifically proposed their HCAI frameworks. These frameworks aim to place human needs, values, wisdom, abilities, and roles at the forefront of AI design, development, and deployment; the ultimate goals are to ensure reliability, safety, and trustworthiness in AI systems while empowering people, amplifying human performance, and supporting human self-efficacy, creativity, and responsibility [1], [2].

The conceptual foundations of HCAI are extensively discussed in recent literature; however, the practices and methodology appear to lag [20], [21], [25], [26], [27], [28]. Although HCAI also emphasizes social issues, such as ethical AI, current HCAI practice has not been systematically practiced from a broad sociotechnical perspective [20], [28], [29], [30], [31]. For instance, researchers have recognized the importance of HCAI in organizational design and best practices to enhance its influence in developing and deploying AI technology [32].

Manuscript received 15 April, xxx. Wei Xu is with the Department of Psychology and Behavioral Sciences and the Center for Psychological Sciences at Zhejiang University, China. Zaifeng Gao is with the Department of Psychology and Behavioral Sciences at Zhejiang University, China. Corresponding authors: Wei Xu, weixu6@yahoo.com; Zaifeng Gao, zaifengg@zju.edu.cn.

Sociotechnical systems (STS) theory has enjoyed seven decades of development and application [33], [34], [35], [36]. Its over-arching philosophy embraces the joint optimization between social subsystems and technical subsystems. It addresses the concerns that technical communities merely focus on technical solutions to address problems. HCAI and STS share the same goal: to develop and use AI to serve humans and society better. From the STS perspective, current HCAI practices face complex challenges arising from the intersection of technology, humans, organizations, and the broader social system (e.g., ethical and legal subsystems) [37], [38]. Thus, STS promises to support HCAI practice toward human-centered AI technology to minimize its adverse impacts on society. We argue that applying STS-based sociotechnical thinking to developing and deploying AI technology will help further enable HCAI in practice and address significant societal challenges.

On the other hand, traditional STS theory has primarily been used to deal with mechanical and non-AI computing systems for many years [33], [39], [40]. AI has brought about new sociotechnical characteristics, contexts, situations, and relationships. For example, the potential human-AI collaboration as a new form of human-machine relationship, AI's cognitive capabilities based on its autonomous agents, unpredictable and unexplainable system output [4], [6], [7]. The emerging sociotechnical characteristics of AI prompt us to reassess and enhance the traditional STS approach. Researchers have initiated investigations into the emerging sociotechnical characteristics brought about by AI and identified the need for enhancing STS theory to support the development and deployment of AI technology so that AI can best serve humans and society [29], [30], [32], [41], [42], [43], [44].

Thus, the critical questions to be answered by this paper are:

*Q1: How should traditional STS theory be extended to represent the emerging sociotechnical characteristics and needs with respect to AI?*

*Q2: How can HCAI practice broaden its scope and impact from the perspective of sociotechnical thinking?*

The contributions of this paper are to extend the traditional STS theory to meet the needs of AI and to further promote current limited HCAI practice by extending its thinking to a sociotechnical perspective; thus, we can maximize the benefits of AI for society, delivering truly human-centered AI systems.

This paper is structured as follows. First, we conducted a literature review of HCAI and STS. We then compared the sociotechnical characteristics of AI technologies with those of non-AI technologies. Based on these findings, we proposed an updated set of STS design principles that capture the key sociotechnical attributes of AI and provide guidance for designing and implementing modern STS in the context of AI. Building on these principles, we introduced a framework for intelligent Sociotechnical Systems (iSTS) to extend traditional STS theory. Informed by the sociotechnical insights of the iSTS framework, we developed a hierarchical HCAI (hHCAI) approach, offering a structured extension of current HCAI practices from a broader sociotechnical perspective. This approach comprehensively addresses gaps in HCAI practices by integrating sociotechnical considerations. Finally, we provide recommendations for further enhancing the iSTS framework and implementing the hHCAI approach. These recommendations aim to mitigate AI's potential negative societal impacts while ensuring the development and use of AI technologies align with human values and societal needs.

## II. Methodology

This paper adopts a qualitative approach to reviewing seminal works in STS and HCAI. A comprehensive *literature review* was conducted, drawing from databases such as Google Scholar, ResearchGate, ACM Digital Library, IEEE Xplore, and other reputable online sources. This extensive search ensured a broad coverage of critical contributions in these domains, establishing a robust foundation for the study.

Two analytical methods were employed based on the literature review. First, a *comparative analysis* identified the sociotechnical characteristics distinguishing AI technologies from non-AI technologies. Second, a *gap analysis* highlighted the limitations in existing STS design principles, revealing areas where they fall short in addressing the complexities of AI technologies. This analysis informed the development of a new set of STS design principles that capture the key sociotechnical attributes of AI and offer practical guidance for designing and implementing modern STS with respect to AI.

A *framework development* method was then applied to create a conceptual framework for intelligent sociotechnical systems (iSTS)**.** This framework was refined iteratively, drawing on seminal works, reassessing the proposed STS design principles, and continuously improving the framework.

Finally, a *case study* method demonstrated the practical implications of the iSTS framework. In this case study, we applied iSTS to current HCAI practices, leading to the proposal of a hierarchical HCAI (hHCAI) approach. This approach addresses the gaps in existing HCAI practices from an STS perspective, offering a structured solution for integrating sociotechnical considerations into AI-driven systems.

## III. Human-Centered AI And Sociotechnical Systems Approaches

### A. Human-centered AI (HCAI)

HCAI is a design philosophy and approach that prioritizes humans in designing, developing, and deploying AI-based intelligent systems. HCAI emphasizes AI being designed, developed, and used to amplify, augment, empower, and enhance human performance rather than harm and replace them [1], [2], [17]. HCAI advocates the creation of AI systems that are deeply aligned with human values, needs, and ethical principles while building advanced AI technologies. The goal of HCAI is to develop human-centered AI technology (including AI-based intelligent systems, tools, and applications) to ensure that AI technology serves humans and society as well as enhances human capabilities rather than harming humans and society.

Over the last several years, researchers have begun to take a human-centered perspective in developing AI technology, such as human-centered explainable AI [23], [45], inclusive design [46], human-centered computing [47], human-compatible AI [48], and human-centered machine learning [49]. For example, Shneiderman [1] and Xu [2] proposed their systematic HCAI frameworks. Specifically, Xu [2] proposed an HCAI framework that systematically addresses the concerns from the aspects of users, technology, and ethics. Recently, Xu et al. [28] proposed an HCAI methodological framework to address the weakness of current HCAI frameworks in guiding practice.

Although HCAI has been heavily discussed in recent literature, current HCAI practice, from a sociotechnical perspective, has not done enough systematic work to address complex challenges that arise from the intersection of technology, humans, and the broader social system [29], [30], [32], [44], [50]. Given that AI systems operate within specific sociotechnical environments and AI may have adverse effects on humans (such as privacy, ethics, and decision-making authority), the negative impacts have prompted researchers to consider the development and use of AI systems from a broader sociotechnical context. For example, Xu et al. [44] argued that HCAI should be applied from the broader perspectives of the ecosystem and sociotechnical environments beyond the current practice that primarily focuses on individual human-AI systems [28]. Thomas et al. [32] argue that HCAI practice should be considered from the perspective of organizational design.

### *B. Sociotechnical Systems (STS) Theory*

Sociotechnical systems (STS) theory has enjoyed around 60 years of development and application [33], [35], [36], [51], [52], [53], [54], [55], [56], [57], [58]. The STS concept originated from the Tavistock Institute of Human Relations during the 1940s and 1950s [33]. Trist and Bamforth [33] explored the relationship between social and technical aspects of work environments, particularly in British coal mines. Based on the impact of the introduction of power machines on work, organization, management, labor relations, employee life, employee families, coal miners associations, and so on, they found that an organization is both a social system and a technical system, and the technical system has an essential impact on the organization and society. This study highlighted how technological changes in coal mining impacted social structures and worker morale, emphasizing the need for joint optimization of social and technical sub-systems. Their key contributions have shaped the development of STS theory and practice, focusing on the interplay between social and technical elements in organizational settings and arguing that neither the social nor the technical subsystem should be optimized in isolation [33].

Figure 1 illustrates the STS approach adapted from [59], [60]. More specifically, a sociotechnical system includes two independent but interdependent social and technical subsystems, among which the social subsystem generally includes people (e.g., knowledge, skills, attitudes, values, and needs), roles, work, culture, and goals involving organizations (e.g., change, structure, decision-making, rewards), and so on. Technical subsystems generally include physical infrastructure, tools, technologies, and processes. STS theory emphasizes the joint optimization between the social and technical subsystems, and the overall system performance relies on design optimization and complementarity between technical and social subsystems. Focusing on one subsystem in the STS to exclude another will decrease the overall performance [33], [35], [55], [57].

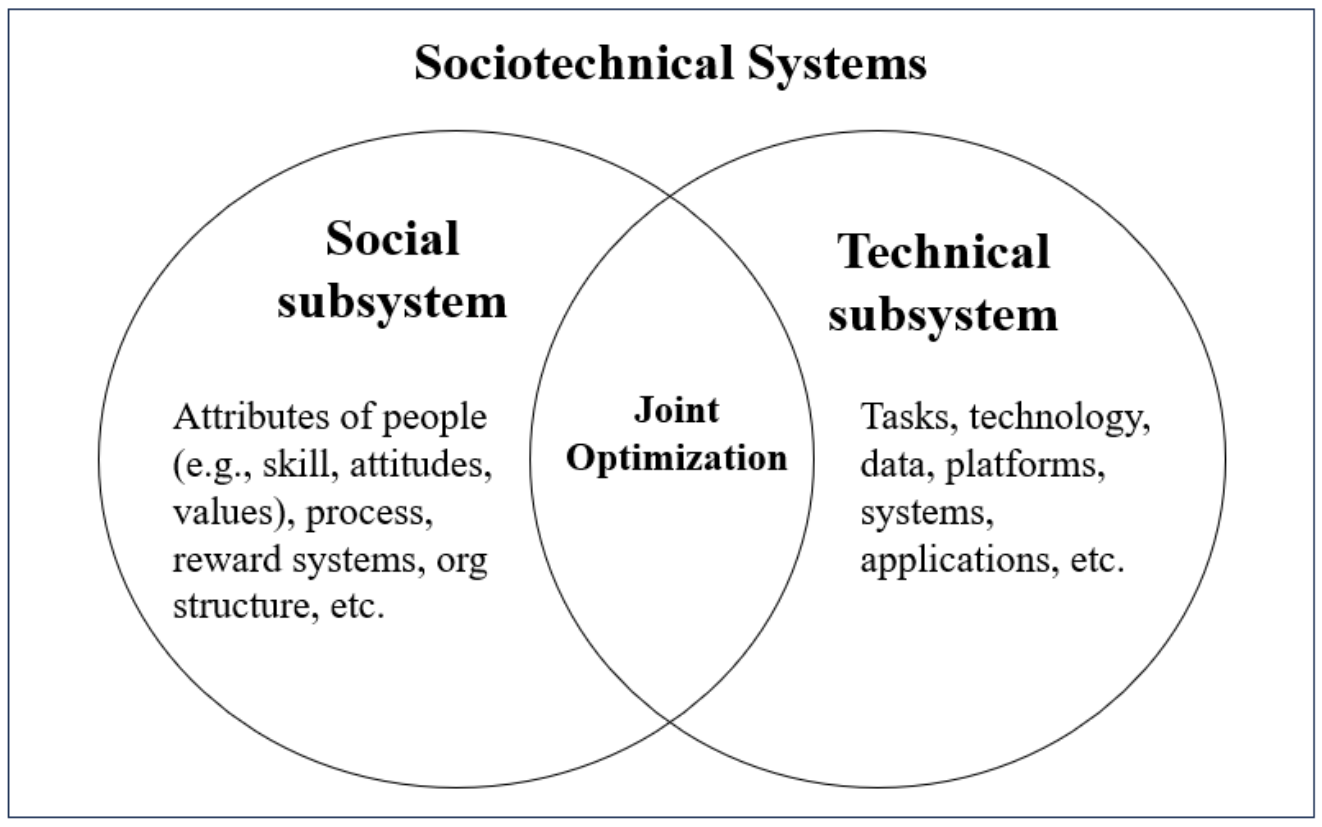

**Fig. 1.** Illustration of the Sociotechnical Systems Theory (adapted from [48], [49])

Introducing new technologies aims to promote knowledge sharing, learning, innovation, optimized design, etc., within organizational and social environments. It intends to help facilitate the easy adoption of new technologies and improve organizational effectiveness. However, new technologies may affect how people work and behave, possibly leading to changes in their values, cognitive structures, work and lifestyles, etc. The joint optimization between technical and social subsystems can help accelerate the adoption of new technologies and improve organizational performance, ultimately improving the quality of human work and life [61].

Modern AI demands ongoing efforts to integrate AI to enhance the interaction between social and technical components, creating an emerging sociotechnical environment in our society [62]. AI tools support complex decision-making processes in organizations, enhancing strategic planning and operational efficiency and improving the accuracy and efficiency of decision-making processes [63]. AI analyzes large datasets to provide insights and recommendations, allowing humans to make more informed decisions [64]. It improves variance control and error management by predicting and identifying real-time errors or issues, allowing immediate corrective actions and reducing system downtime [65]. AI is also used for adaptive system design by adjusting to changing conditions, providing flexibility and scalability in system design and operation [66]. Additionally, AI technologies automate routine tasks, enabling human workers to focus on more creative and strategic activities [67]. AI enhances the resilience and robustness of sociotechnical systems, ensuring they can withstand and adapt to disruptions [68].

Researchers have begun to examine the social implications of AI from a sociotechnical perspective, including impacts on employment and workforce dynamics [69], the intertwined relationship between AI and societal inequalities [70], the moral significance of AI within sociotechnical frameworks [71], the integration of AI within organizational settings [30], evaluating the safety of generative AI and the importance of explainability in AI systems from a sociotechnical perspective [72], [73]. There is a growing emphasis on integrating ethical principles into AI systems to ensure fairness, transparency, and accountability [29], [30]. These efforts align with principles that promote trust and acceptance of AI deployment in sociotechnical systems [74], [75]. Beyond influencing the design and use of AI technology, STS theory has also impacted organizational design, e.g., the design of jobs and ways of organizing work [32], [34]. Thus, AI has brought opportunities and transformative changes to the STS approach. While progress has been made, challenges remain in fully realizing AI's potential as sociotechnical systems. The STS approach needs to evolve further to enable it to provide better support for the development and deployment of AI technology.

### C. Implications of STS to HCAI

Developing and using new technologies must occur within a sociotechnical environment. AI-based intelligent systems have added complexity to this environment with their unique characteristics such as autonomy, cognitive capabilities, machine learning and evolution, explainability, and potential for system bias. On the one hand, increasingly intelligent systems influence societal aspects like ethics and social culture. On the other hand, current research on intelligent systems from the perspective of STS remains insufficient [29], [76].

Integrating HCAI principles into the development of intelligent systems while ensuring technical robustness and efficiency can be challenging [1], [2], [3]. STS theory advocates a joint optimization between social and technical subsystems, which may sometimes be in conflict. Also, intelligent systems can have unintended consequences due to their unique characteristics, especially in complex social settings [77]. Anticipating and mitigating these consequences while aligning with human-centered values is a significant challenge [1], [2], [3]. For long-term sustainability and from a sociotechnical perspective, it is essential to ensure that intelligent systems are sustainable in the long term, both technically and socially [1], [3], [78].

HCAI considers human, social, organizational, and technical factors in designing, developing, and deploying intelligent systems. The outcome of applying HCAI is a better understanding of how human, social, and organizational factors affect how intelligent systems are designed, developed, and deployed. In recognition of STS value, researchers have called for sociotechnical thinking for HCAI [1], [29], [30], [32], [44], [79]. Some researchers have begun shifting their research on intelligent systems from purely technical to a comprehensive approach encompassing social, organizational, and other factors [30], such as human-centered explainable AI from an STS perspective [45]. Naikar et al. [80] proposed designing human-AI systems for complex environments from a distributed, joint, and self-organizing perspective of STS. Xu et al. [28] presented an HCAI concept that intends to extend the current focus of HCAI from the emphasis on individual human-AI systems to a broad sociotechnical perspective.

Organization design is a core component of STS. In current HCAI practices, although organizational alignment is considered an indispensable component of sociotechnical systems, HCAI approaches do not systematically consider the organization involved [32]. To address the issues, Herrmann et al. [32] have attempted to fill the meso-level gap between keeping individual humans and keeping society in the loop and proposed the concept of "keeping the organization in the loop (KOITL)" as a new metaphor [32], [81]. They present a systematic view of how AI and organizational practices must be intertwined as a prerequisite and continuous context of HCAI. Thus, HCAI practice must be accompanied by a specification of the organizational practices that precondition its success .

Based on the review of literature presented above, there is a strong connection between the HCAI and STS approaches when it comes to designing, developing, and deploying AI systems. Both approaches have a common goal of improving human performance and well-being. HCAI and STS also share similar methods. For example, sociotechnical principles emphasize the importance of involving users in designing and implementing AI systems to ensure that AI tools meet user needs. STS theory emphasizes the importance of people and organizations when developing and introducing new technology from a broad sociotechnical perspective. Therefore, it is worthwhile for us to reassess STS theory and its implications for HCAI practice from a broader sociotechnical perspective, further enabling HCAI in practice and maximizing its contributions.

## IV. Emerging Challenges to Traditional STS Theory

### A. New sociotechnical characteristics in AI

STS theory was developed in the 1950s, and since then, the focus of STS research has been too narrow, there are new contexts and problems that could benefit substantially from STS thinking [34]. There are challenges to applying traditional STS theory in AI, which is driven by the unique characteristics of AI technology. Earlier STS practices designed social systems based on fixed technology and fixed processes to improve worker productivity and controllability over their work [33]. Systems such as the Internet of Things and smart transportation will surely be realized in complex sociotechnical ecosystems where various factors are interdependent; multiple factors in the ecosystem will also affect their development and use.

Also, AI-based technical subsystems have far exceeded the scope of non-AI-based digital computing systems. For example, AI technology and its autonomous characteristics promote the role of machines, from simple tools assisting human work to potential collaborators with humans, thus forming a new form of human-machine relationship facilitating human-AI collaboration [6], [78]. As discussed earlier, while AI has brought benefits to humans, it may cause many issues. From a broader sociotechnical perspective, the development and use of AI systems have also triggered a series of new issues in the areas of AI ethics, user privacy, human skill retention, organization design, decision-making authority, understanding and acceptability of AI, and so on. These emerging challenges require a new way of thinking to consider how to solve the new problems more effectively for humans and society from a broader sociotechnical perspective. Table 1 compares sociotechnical characteristics between the non-AI computing and AI technologies [3], [44], [78], [83], [84], [85], [86]. As shown in Table 1, AI technology presents a series of emerging sociotechnical characteristics and challenges. AI systems will promote closer interaction and collaboration between humans, machines, organizations, and society. While traditional STS theory has made an impact, we need to evolve and extend its reach to meet the challenges of AI so that we can effectively solve new problems brought about by AI technologies, maximizing AI's advantages and avoiding its negative impacts on society.

*B. Updating STS design principles to meet the needs of AI*

STS design principles are propositions that serve as the foundation for a system of belief or behavior [87]. These design principles are crucial for creating systems that effectively achieve organizational goals and support the people who operate within them. Principles are "guides to critical evaluation of design alternatives making clear some of the differences between the sociotechnical systems approach and traditional job design" [88]. Good STS design principles should be succinct and clear, clarifying the desired outcomes and who should act on them [87]. They help achieve the goals of sociotechnical systems by balancing technical advancements with human needs, fostering an adaptable, resilient environment conducive to continuous improvement and innovation.

STS design principles have been derived from many STS research and practices. STS design principles are essential for enriching STS theory and practice [87]. Over the course of STS development, sets of principles have been developed and revised [33], [39], [40], [82], [87], [88], [89], [90], [91], [92], [93]. These design principles have helped guide how new technology may be used and integrated with social systems [94]. STS theory has evolved significantly since its inception, with contributions from multiple scholars emphasizing the need for integrated design approaches that consider both social and technical dimensions. These seminal works have laid the groundwork for STS practices, influencing how organizations design and manage complex systems.

STS design principles evolve over time. A lot has changed since 1946 when the STS theory was proposed and the first sociotechnical principles were developed, but the need to update STS design principles remains [87]. For example, Cherns [82] proposed design principles in 1976 and revised the principles in 1987 [83]. In the revision, more attention was given to the needs of the organization as a society and new principles (e.g., power and authority, transitional organization) [83]. Recently, Imanghaliyeva et al. [87] took a systematic review, grouped design principles by similarity, and synthesized them into an updated set of principles. As a result of the systematic review, they created a list of 20 STS design principles that were extracted from the broader literature [87]. However, Mumford [36] argued that although technology may change, some basic principles remain the same, such as, humans (employees) must be given high priority.

The fortunes of sociotechnical theory have ebbed and flowed over the past seventy years, but the value of sociotechnical principles has remained [87]. For example, the "joint optimization" design principle aims to optimize social and technical components, leading to higher productivity, quality, and reliability. This design principle has been adopted and practiced since the inception of STS theory [33], [36], [39], [55], [62], [82], [87], [88], [89], [90], [95] .

We argue that STS design principles should continue evolving as AI technologies emerge. Based on the emerging sociotechnical characteristics, challenges, and needs brought about by AI technology, we explored updating the existing design principles with additional principles (see Table 2). As shown in Table 2, our initially proposed design principles are defined based on the emerging sociotechnical characteristics (Table 1) and the previous seminal work of the STS community to address the sociotechnical environment in AI. Specifically, as discussed earlier, the "joint optimization" design principle has been revised to the "human-centered joint optimization" design principle to reflect the increased importance of human-centeredness in AI. This revision emphasizes the human-centered characteristics of AI, as shown in Table 2 (column 4).

Furthermore, the traditional human-machine (non-AI technology) relationship is transitioning to a human-AI relationship within the AI-based sociotechnical environment. This is represented by various forms, such as human-AI collaboration, human-AI complementarity, human-AI decision-making sharing, and human-AI co-learning/co-evolving. Table 2 shows that the updated STS design principles highlight human leadership in human-AI collaboration, human roles/functions in human-AI complementarity, human authority in shared decision-making, and human guidance in human-AI co-learning/evolving. We argue that these human-centered STS design principles will enable AI technology to empower humans. The other three design principles listed in Table 2—(1) 'flexibility and adaptation,' (2) 'ethical and responsible design,' and (3) 'transparency and explainability'— address the new challenges of AI technology and the emerging sociotechnical characteristics, as highlighted in Table 1 [23], [24], [45], [96].

TABLE I
COMPARISON OF THE SOCIOTECHNICAL CHARACTERISTICS BETWEEN NON-AI AND AI TECHNOLOGIES
(SOURCES: [3], [44], [78], [83], [84 ], [85], [86])

| **Items** | **Sociotechnical characteristics of non-AI technology** | **Sociotechnical characteristics of AI technology** |
|---|---|---|
| Design philosophy | Human-centered | Human-centered |
| Machine's role | Tools supporting human tasks | Plus: potential collaborators with humans |
| Human-machine relationship | Human-computer interaction | Plus: potential human-AI collaboration |
| User needs | Experience, safety, functionality, etc. | Plus: ethics, skill growth, decision-making, etc. |
| Quality of user interface | Usability, functionality | Plus: AI explainability and understandability |
| Decision-making | Human only | Shared decision-making with human authority |
| Learning ability | Human only | Plus: human-AI co-learning and co-evolving |
| Cognitive capability | Human only | Plus: AI machine also has (e.g., sensing, reasoning) |
| System output | Expected as design | Plus: potentially biased, unpredictable |
| Scope of design | Internal organizations, static systems | Plus: ecosystem (across organizations), social environments |
| Ecosystem | Limited | Intelligent ecosystems (smart traffic, city, etc.) |
| System dynamics | Static and fixed process, technology | Changing technical systems driven by AI |
| Design goals | Optimization between technical and social subsystems | Plus: ability to cope with uncertain changes inside/outside the system |
| Organizational goals | High performance, control over change, a stable system | Innovation, agility, and adaptability to respond to dynamic changes |
| Organizational needs | Work system design (technology, role, process, skill, etc.) | Plus: human led collaboration with AI, shared decision-making |
| Complexity and openness | Relatively independent and closed systems. | Fuzzy boundaries, system evolving and unpredictable |

TABLE 2
PROPOSED STS DESIGN PRINCIPLES (SOURCE: ORIGINAL CONTRIBUTION BY AUTHORS)

| ***Design principle*** | **Proposition** | **Emerging sociotechnical characteristics of AI** | **Human-centered characteristics** |
|---|---|---|---|
| Human-centered joint optimization | Optimize both social and AI subsystems to empower human capabilities and enhance overall performance. | AI may negatively impact humans and society if humans are not prioritized in design, development, and deployment. | Human needs (e.g., trust, ethics), roles, values, and authority. |
| Human-led collaboration with AI | Promote human-AI collaboration and ensure humans are leaders on the human-AI teams. | AI may collaborate with humans beyond interaction with humans. | Human leadership. |
| Human-led decision-making | Enhance decision-making by leveraging the strengths of humans and AI while maintaining human's ultimate authority. | AI and humans share decision-making by leveraging both strengths, resulting in challenges of collaborative processes and roles. | Human ultimate authority. |
| Human-AI complementarity | Achieve more powerful system intelligence while maintaining the key roles of humans by complementing human and machine intelligence. | AI has strengths in many aspects; humans and AI share many roles and tasks. | Human roles and functions. |
| Human-guided co-learning and co-evolving | Facilitate the co-learning/co-evolving between humans and AI with human guidance, adaptively handling dynamic and complex operations. | Humans and AI can co-learn and evolve. Their roles and authority become crucial in guiding and supervising human-AI systems. | Human oversight and governance. |
| Flexibility and adaptability | Design systems to adapt to AI advancements, organizational changes, and scenarios that design cannot predict. | AI has unique autonomous characteristics that can handle some unexpected events that are not predictable by design but may cause uncertainty. | AI systems adapt to humans, not vice versa. |
| Ethical and responsible design | Ensure that ethical considerations are at the forefront of system design. | AI generates emerging user needs (e.g., privacy, fairness). | Emerging human needs. |
| Transparency and explainability | Ensure system outputs are explainable and interpretable, enabling users to understand how decisions are made. | AI has a "black box" effect that causes transparency and explainability issues. | Human trust and understanding. |
| Systemic design | Facilitate multiple-level design across individual human-AI systems, organizations, ecosystems, and social systems. | The sociotechnical environment of AI is dynamic and open, with more uncertainty and blurred boundaries across systems. | Human needs, values, and well-being across all levels of interaction. |

Thus, we believe that these updated STS design principles should help us design human-centered AI solutions in AI-based sociotechnical systems, achieving the goals of the STS approach by balancing AI advancements with human needs and emerging sociotechnical characteristics brought about by AI. Although these design principles are to be validated and refined through future work, they may help us enrich existing STS theory with respect to AI. The next section will elaborate on how these principles are applied in developing our iSTS framework.

*C. Previous seminal work for updating traditional STS theory*

The research and application of STS theory have been carried out for decades [97]. Over time, its emphasis has shifted from an early focus on heavy industry [33] to advanced manufacturing technologies [40] and then to office-based workplace services [39], [41], [42]. Although the applications have evolved, the underlying philosophy of STS has remained unchanged [89]. With the rise of AI technology, researchers are calling for advances in STS theory to address the unique challenges and opportunities presented in the context of AI.

Below, we highlight key insights from recent research outcomes. These previous studies have provided valuable. insights into updating the traditional STS theory.

- Michael et al. [62] argue that cyberharms persist at *individual (micro), organizational (meso), and societal* (macro) level. They identified 10 gaps and built a research roadmap to deploy a socio-technical approach to AI in cybersecurity.
- Davis et al. [89] argue that we need to explore *new opportunities* to apply STS thinking to new problems, test existing principles' adequacy, and identify where STS can contribute to new fields. Although their proposal did not specifically address AI, they believe the focus of STS has been too narrow.
- Huang et al. [98] used the notion of system levels to illustrate an STS concept that meaningfully interconnects *multiple levels* of consideration. The upper two levels (group/community and personal/individual) are social, while the lower two (information/software and physical/hardware) are technical.
- Steghofer et al. [99] believe that the next generation of STS should be *based on AI technology*. They emphasize dealing with the structure of adaptive next-generation intelligent systems, transitioning from the paradigm of the literate user to collaboration between the user and AI systems.
- Dalpiaz [100] proposed an *adaptive STS* system architecture that enables self-reconstruction through a monitoring-diagnosis-coordination cycle. This architecture for self-reconfigurable STSs is based on goal-oriented requirements, considering the dynamic, unpredictable, and weakly controllable characteristics of STS in AI.

- Thomas et al. [32] argue that integrating human and machine intelligence is achievable only if human organizations—not just individual human workers—are kept "in the loop." They suggest a systematic framework for "*keeping the organization in the loop*" based on interacting organizational practices.
- Makarius et al. [78] developed an integrated model based on STS theory that combines AI novelty and scope dimensions. They adopt an organizational socialization approach for the process of *integrating AI into organizations*.
- Winby et al. [85] proposed a design method for digital STS. Although the proposal does not fully consider the emerging characteristics of AI, they argue that existing STS frameworks do not adequately address the *new reality* where both technical and social elements of the full ecosystem need to be designed. STS work is no longer confined within a bounded organization, and individual organizations can no longer be the sole design focus.
- Taddeo et al. [43] emphasize broader *sociotechnical ecosystem* considerations for achieving system robustness for AI in cybersecurity. Their approach highlights people, processes, and technology at various levels relevant to individuals, organizations, and society. They suggest future research requires an interdisciplinary approach.
- Rahwan [81] proposes a conceptual framework for keeping "*society-in-the-loop*" for the regulation and governance mechanisms of AI systems. This framework combines the "human-in-the-loop" control paradigm with mechanisms addressing the values of stakeholders affected by AI.
- Xu et al. [44] proposed a new research paradigm of *human-AI joint cognitive ecosystems* to facilitate HCAI practice. An intelligent ecosystem with multiple human-AI systems can be represented as a human-AI joint cognitive ecosystem. The overall system performance depends on the optimization across multiple human-AI systems from a broad ecosystem and social perspective.
- Pasmore et al. [84] focused on the 'next generation' sociotechnical design and proposed a framework for sociotechnical systems design specifically for future organizations. The framework includes *multi-level design*: strategic design, operating system design, and work design. The goal is the balanced optimization of the ecosystem, organization, technical system, and social system.
- Baxter et al. [35] analyzed why STS designs (STSD) are not widely practiced. Some failings of STSD can be attributed to the *multidisciplinary nature* of system design. They identified the need for interdisciplinary teams in the process and proposed a framework of Sociotechnical Systems Engineering based on research in work design and cognitive systems engineering, aiming to bridge the gap between organizational change and systems development.

## V. Toward Intelligent Sociotechnical Systems (iSTS)

In the previous two sections, we defined the emerging sociotechnical characteristics (Table 1) and the emerging STS design principles (Table 2) with respect to AI. Although these studies have addressed some gaps in various ways, no comprehensive STS concept has systematically addressed the emerging sociotechnical characteristics and design principles. There is a clear need to extend the traditional STS theory to support the development and deployment of AI technology from a sociotechnical perspective. Based on our initial concept of intelligent sociotechnical systems (iSTS) that extended the traditional STS theory [44], [79], we further extend the initial iSTS concept to an iSTS framework, as illustrated in Figure 2.

Compared to the traditional STS theory (Figure 1), the iSTS framework introduces several distinctive features. While it retains core elements of STS, such as the principle of "joint optimization"—which emphasizes balancing performance between technical and social subsystems—the iSTS framework advances this concept into "human-centered joint optimization." This evolution addresses emerging human-centered needs in AI, ensuring that optimization considers not only system performance but also human well-being and values.

The iSTS framework is built on a revised set of STS design principles (see Figure 2, gray box in the center), adapted to the unique characteristics of AI technologies. The technical (AI) subsystem highlights essential AI-specific features, such as agents, autonomy, unique machine behavior, and cognitive capabilities. On the other hand, the social subsystem emphasizes critical aspects required by AI-integrated systems, including ethical considerations, human authority in decision-making, empowering human abilities rather than replacing them, and the need for substantial work system redesign.

Also, the iSTS framework introduces human-centered joint optimization across four levels: individual human-AI systems, organizations, ecosystems, and social systems. This hierarchical approach demands multi-level design and integration across these levels, ensuring that human-centered principles are embedded throughout the system. In the following sections, the new features of the iSTS framework are characterized as follows compared to traditional STS theory.

More specifically, the iSTS framework demonstrates the following differentiated features as compared to traditional STS theory.

### A. Human-centered approach

Guided by the updated STS design principles, iSTS adopts a human-centered approach. As outlined in Table 2, each design principle represents and realizes this approach. Through design principles such as human-centered joint optimization, human-led collaboration, human-led decision-making, and human-guided co-learning/evolving, iSTS aims to achieve human-centered characteristics in AI-based sociotechnical systems. Moreover, as illustrated in Figure 2, the human-centered approach should be practiced across the four levels of human-AI systems, organizations, ecosystems, and social systems [95], [98].

### B. A hierarchical approach

Figure 2 illustrates the hierarchical structure of iSTS across individual human-AI systems, organizations, ecosystems, and social systems. Unlike traditional STS focusing on individual human-machine systems and internal organizations, iSTS goes beyond organization boundaries with systematic thinking for such complex and open systems as have emerged today [43],

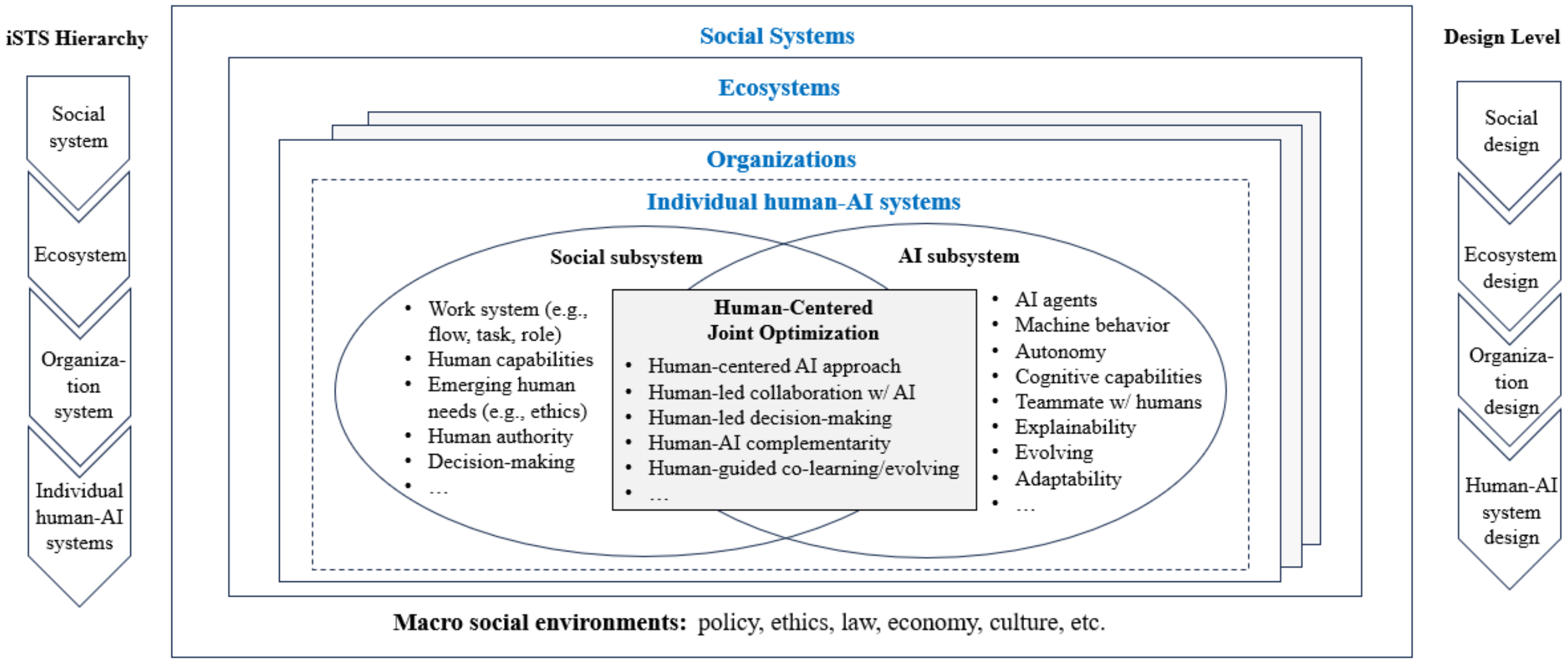

**Fig. 2** Intelligent Sociotechnical Systems (iSTS) (adapted from [44], [79])

[44], [81], [85], [37], [38], [101]. This implies that designing, developing, and deploying AI systems should not only focus on a "point" solution (i.e., focusing on individual human-AI systems and organizations) but also consider the optimization among multiple human-AI systems within ecosystems. Within such ecosystems, humans and intelligent agents work together in an interconnected fashion using technological platforms to achieve a shared purpose as the ecosystem evolves as quickly as the external environment demands [95]. iSTS argues that the optimal design of individual human-AI systems within organizations cannot guarantee the optimal performance and safety of the entire intelligent ecosystem. The optimization of an intelligent system must systematically consider the optimization of other systems within its ecosystem. Furthermore, the systematic thinking of iSTS continues to the perspective of macro social systems, seeking an end-to-end approach at a social level (i.e., focusing on the entire sociotechnical systems as a whole). This design thinking enables us to explore system approaches to achieving optimal design, development, and deployment of intelligent systems through the effective integration between technology and non-technical factors from a systematic perspective [34].

### *C. A multi-level design approach*

Figure 2 illustrates that iSTS proposes a multi-level design approach through multidisciplinary collaboration, including individual human-AI system design, organization design, ecosystem design, and social system design. Designing, developing, and deploying an AI system is no longer a technical project but an STS project involving collaboration across multiple disciplines. The goal of all four levels of design is to achieve a balanced optimization of the technical system, organization, ecosystem, and social system. In the traditional STS approach, the goal typically was to design the social system around a fixed technical system to satisfy human needs. In iSTS, the goal of optimization is predicated on the dynamics of the entire organizational, ecosystem, and social environments. As the environment changes, the design of the four components will evolve and need to be aligned for a better fit between the subsystems and the environment, thereby increasing sustainability [95]. The success of iSTS needs to be measured across the design levels. Thus, although organization design remains one core component from the sociotechnical perspective, iSTS extends the scope of the traditional STS approach beyond "organization design," driving comprehensive solutions for AI systems across all levels.

### *D. Human-led collaboration with AI*

Unlike traditional STS theory, which focuses on the interaction-based relationship between humans and non-AI technical systems, iSTS considers an AI system as a potential collaboration of the human-AI system, not just a simple tool, as presented in traditional STS. Such a shift in design helps explore ways to improve the overall system performance by optimizing human-machine collaboration [6]. The overall system performance depends on the performance of individual parts and the human-machine intelligence complementarity supporting mutual collaboration [99]. The interdependent human-AI system in iSTS shares its goals. Mutual trust, shared information, shared situation awareness, and shared decision-making and control between humans and AI agents are essential to iSTS while maintaining humans as the ultimate authority. Also, characterizing humans and AI agents in human-AI systems as potential collaborators helps draw on mature human-human teamwork theories and methods to develop theories and methods for human-AI collaboration. This work will help study, model, design, build, and verify the collaborative relationship between humans and AI agents to improve overall system performance.

### E. Human-AI complementarity

Human-AI complementarity refers to the synergy between human intelligence and AI, where each complements the other's strengths and mitigates weaknesses. This creates a unique opportunity, which we have not seen prior to the introduction of AI, so we can leverage the unique capabilities of humans and AI to achieve outcomes that neither could accomplish alone. Human-AI hybrid intelligence involves creating structures, processes, and cultures that optimize the synergy between human capabilities and AI. iSTS calls for redesigning human and AI systems as a new type of work system and optimizing the function/task distribution between them based on the complementary advantages between humans and AI, including their roles, workflow, operating environment, organization structure, decision-making authority, etc. From an organizational design perspective, organizations can enhance their performance, innovation, and adaptability in a rapidly changing technological landscape by leveraging the strengths of both humans and AI. Integrating human-AI hybrid intelligence involves structuring the organization to optimize the collaboration between human employees and AI systems. This integration aims to enhance productivity, innovation, and decision-making while maintaining a balanced, ethical, and adaptive organizational culture. This approach requires careful consideration of ethical, social, and interactional factors to ensure that the resulting systems are beneficial and trustworthy.

### F. Human-AI co-learning and co-evolving

In contrast to traditional non-AI computing technology, intelligent machine agents with autonomous capabilities are unique resources demanded by AI. Just as any ecosystem will learn and evolve, iSTS emphasizes the continuous learning, evolution, and optimization of machine agents within intelligent ecosystems [101]. The complex relationships between social and intelligent technical subsystems span the traditional boundaries between humans, machines, organizations, and so on, enabling dynamic interactions and collaboration between AI agents and human agents (at the individual, social, and organizational levels). AI agents facilitate dynamic interaction and collaboration, which in the short term adjusts the behavior of the AI agent itself (based on machine learning algorithms, etc.) and leads to long-term patterns of human use and expectations [102]. Also, the social and technical subsystems contain different types and levels of human and AI autonomy, making learning growth, flexibility, and adaptive capabilities across human-AI systems. Therefore, the design, development, and deployment of AI systems must consider the emerging features of human-AI co-learning and co-evolving through cross-task knowledge transfer, self-organization, and adaptive collaboration so that the ecosystem can coordinate its various components to adapt to dynamic and complex operating scenarios, ultimately improving the overall learning, evolution, and collaborative capabilities of the ecosystem [28], [102].

### G. Open ecosystems

Before widespread AI, the analysis unit in traditional STS theory was usually a bounded part of an organization and was relatively independent. Today, the design and development of various types of intelligent systems, such as the Internet of Things and intelligent transportation, exist in a complex and interdependent sociotechnical ecosystem [101]. The differences between iSTS in modern AI systems and traditional STS lie in the existence of AI agents. In a dynamically developing social environment, the presence of autonomous AI agents will dramatically increase, resulting in a higher degree of uncertainty and unpredictability of AI systems compared to non-AI environments. The cognitive learning of AI technology, uncertainty in output, and other autonomous characteristics bring dynamic and blurred boundaries across systems. These open features bring innovative design opportunities to AI systems but also bring challenges to system design. Therefore, the design, development, and deployment of AI systems need to be considered from the perspective of an open, human, technological, social, and organizational ecosystem.

Finally, we consider the proposed iSTS framework as a starting point for extending the traditional STS theory and addressing crucial issues in AI. The iSTS framework is grounded in emerging sociotechnical characteristics, updated STS design principles, and insights from previous work.

## VI. A Sociotechnically-based Hierarchical HCAI Approach

As an application of the proposed iSTS framework, we integrate it into current HCAI practices, demonstrating its value in advancing HCAI.

### A. Application of iSTS to HCAI

The iSTS framework offers insights into HCAI practice and can help address the current challenges. HCAI and iSTS approaches share the human-centered design philosophy. William Howell [103], past president of the Human Factors and Ergonomics Society, presented a “shared design philosophy” model; the human factors discipline needs to share the human-centered design philosophy with other disciplines instead of exclusively possessing the design philosophy. Over the past 40 years, the human factors community has integrated its approaches and professional talents into the practices of the fields of human-computer interaction and user experience; the growth of these emerging fields fully embodies the shared design philosophy model. With respect to AI, HCAI is the further practice of this model [2], [3]. Collaborations between STS and HCAI professionals will help further propagate the human-centered approach and solve more practical human and social problems.

iSTS fosters a hierarchical approach across the four levels of assessing human interactions with AI systems, ranging from individual human-AI systems, organizations, ecosystems, and social systems. Such a hierarchical approach implies that designing, developing, and deploying AI systems need to be performed systematically with a broader perspective. For example, while current HCAI practice emphasizes ethical considerations and considers human needs, iSTS encourages a broader view of ethics, extending beyond the individual to assess the impact of AI design and use in the context of society as a whole. By considering the various social, cultural, organizational, and environmental factors, iSTS can guide the development and use of AI systems more attuned to diverse human needs and values.

TABLE 3
IMPLICATIONS OF iSTS TO HCAI PRACTICES
(SOURCE: ORIGINAL CONTRIBUTION BY AUTHORS)

| **Aspects of iSTS** | **Characteristics of iSTS** | **Implications to HCAI** |
|---|---|---|
| Design philosophy | Human-centered AI. | iSTS and HCAI share the same design philosophy. |
| Design goal | Optimization between technical and non-technical subsystems for maximizing AI benefits and empowering humans. | iSTS and HCAI share the same design goal. |
| Approach | A hierarchical approach across the hierarchy of individual systems, organizations, ecosystems, and social systems. | Take a hierarchical approach by focusing on the perspectives of the organization, ecosystem, and social environments beyond individual human-AI systems. |
| Design scope | Multiple levels of design across human interaction design with individual AI system, organizational design, ecosystem design, and social system design. | Conduct multiple-level design across individual human-AI systems, organizations, ecosystems, and social systems. |
| Human-machine relationship | Interaction and human-led collaboration between humans and AI agents. | Design, develop, and deploy the collaborative relationship between the two agents to improve system performance. |
| Organizational needs and goals | Optimized organization and work system design (e.g., role, flow, decision-making) for business values. | Optimize organization and work system for AI acceptance, AI-empowered productivity gain, HCAI culture, and so on. |
| Collaboration | Interdisciplinary collaboration across all levels of design. | Develop interdisciplinary HCAI methodology (e.g., methods, tools, processes). |

Table 3 highlights the implications of iSTS for HCAI practice. Therefore, iSTS provides sociotechnical thinking that can further advance HCAI in practice from a broader perspective. Both approaches can complement each other in many ways, and collaboration between them is crucial for developing truly human-centered AI systems.

### *B. A paradigmatic extension: A sociotechnically-based hierarchical HCAI (hHCAI) approach*

Over the years, HCAI has gained significant momentum and has been extensively discussed in literature. While HCAI addresses AI's societal impacts, including ethical issues like user privacy and fairness, current practices primarily focus on individual human-AI systems, such as explainable AI and human-AI interaction design [1], [3], [23], [30], [32], [44].

More specifically, although Taddeo et al. [35] did not explicitly propose an extended HCAI framework, they emphasized that considering a broader sociotechnical ecosystem is crucial to achieving system robustness for AI in cybersecurity. Thomas et al. [32] take the current HCAI practice AI one step further and argue that integrating humans and machine intelligence is achievable only if organizations—not just individual human workers—are kept "in the loop." They suggest a systematic framework of "keeping the organization in the loop" based on interacting organizational practices. Some researchers have begun shifting their research on AI systems from purely technical to a comprehensive approach encompassing social, organizational, and other factors. For example, Stahl et al. [29] propose a novel approach to AI ethics from an ecosystem perspective. Asatiani [30] presents an approach for implementing AI accountably and safely in organizational settings, which paves the way for more responsible and accountable AI in organizations. Schoenherr et al. [23] highlight the importance of addressing explainability and accuracy in AI through an HCAI approach. They encourage people to take a broad view, integrating social and cognitive insights to build trustworthy systems. Xu et al. [44] also argue that HCAI should be practiced from the broader perspectives of the ecosystem and sociotechnical environments beyond the current focus on individual human-AI systems.

However, the current work of AI systems from the perspective of STS is far from sufficient [29]. Based on iSTS, we move one step further and argue that the scope of current HCAI practice needs a paradigmatic extension; that is, HCAI practice needs to adopt a more systematic approach by considering all levels of design, development, and deployment of AI systems. Specifically, we propose a sociotechnically-based hHCAI approach (Figure 3). This illustrates the hHCAI approach, which is currently practiced in the context of individual humans or groups interacting with AI systems (i.e., the focus of current practice), should be extended to the context of organizations, ecosystems, and macrosocial systems.

The hHCAI approach highlights the following new six characteristics beyond current HCAI practice.

1) **A paradigmatic extension**
   Beyond the scope of current HCAI practice that primarily focuses on individual human-AI systems, the hHCAI approach extends the scope of HCAI practice to the perspectives of organizations, ecosystems, and social systems [44].

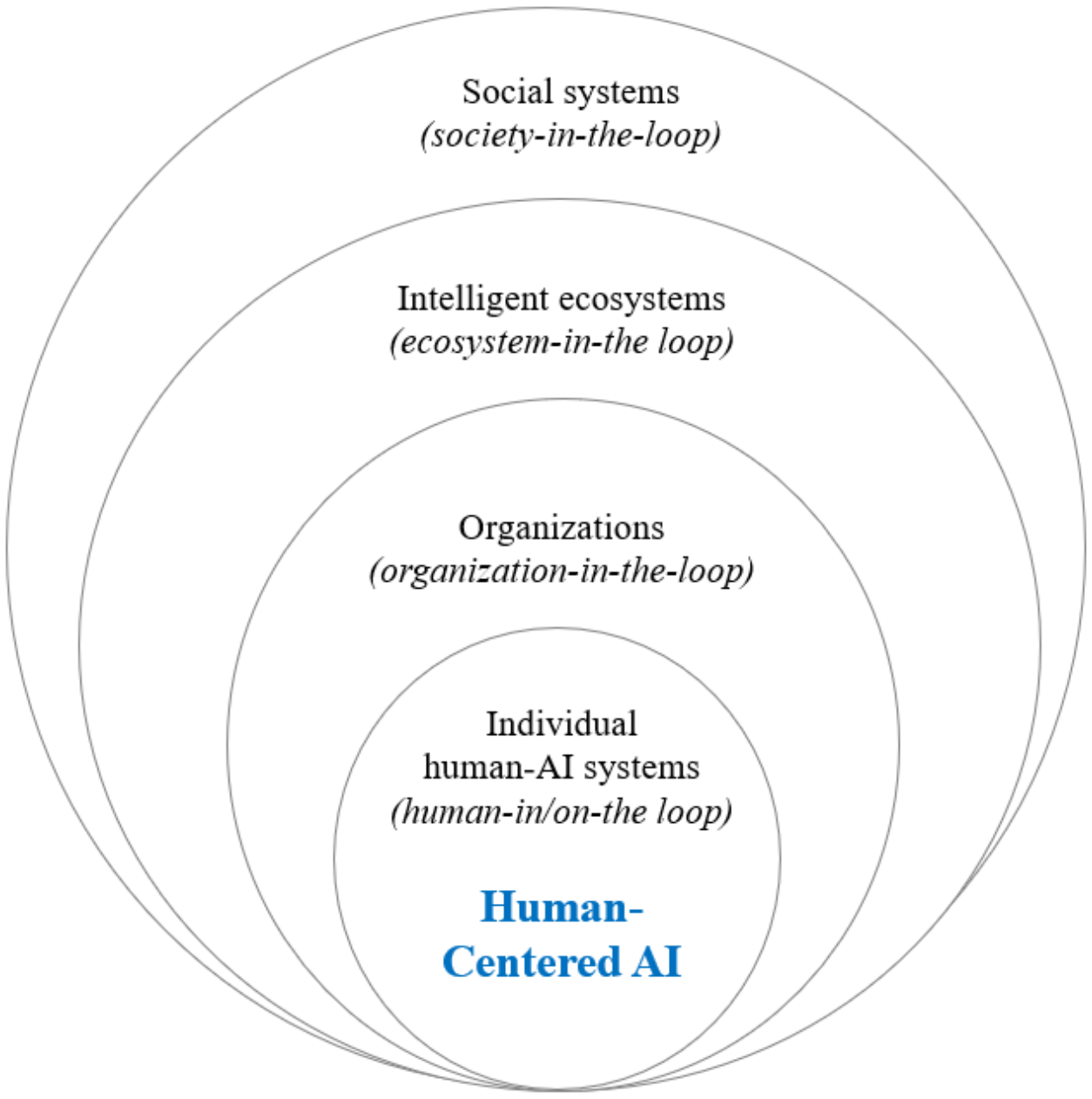

**Fig.3** A sociotechnically-based hierarchical HCAI (hHCAI) approach (Source: Original Contribution by Authors)

2) **Complementary design metaphors**
The current HCAI practice puts more focus on *individual* humans or groups in the optimization of human-centered AI systems, such as the human-in-the-loop and human-on-the-loop design in a narrow context. The hHCAI approach goes beyond that focus to achieve HCAI goals by keeping organizations, ecosystems, and society in the loop of designing, developing, and deploying human-centered AI systems [84]. This approach leads to emerging design metaphors complementing current HCAI design: organization-in-the-loop, ecosystem-in-the-loop, and society-in-the-loop.
3) **Keeping organization-in-the-loop design**
The hHCAI approach calls for adopting the *organization-in-the-loop* design metaphor for organizational design when practicing HCAI within organizations [32], [78]. This helps design and structure an organization to maximize the potential of AI while keeping workers at the design center and focusing on their needs and capabilities. Thus, the organization can effectively integrate HCAI into its structures, operations, and organizational learning and adaptation to AI, eventually enhancing productivity and maximizing the values of AI.
4) **Keeping ecosystem-in-the-loop**
The hHCAI approach requires the *ecosystem-in-the-loop* design metaphor for ecosystem design in HCAI practice. Such an approach involves creating an environment where technology, human values, and organizational structures work in harmony across the boundaries of individual organizations within an ecosystem [43], [85]. For example, this approach will establish systems for continuous learning and adaptation, where AI systems can evolve based on user feedback and changing societal norms; it involves activities such as data privacy and security in the AI ecosystem, partnerships and stakeholder engagement, and regulatory compliance [14], [62], [101]. The goal is to ensure the optimization of multiple AI systems toward human-centered AI ecosystems.
5) **Keeping society-in-the-loop design**
The hHCAI approach requires HCAI practice to keep *society-in-the-loop* of designing, developing, and deploying AI systems by taking a systematic approach to embed societal values [44], [81]. It emphasizes the importance of designing, developing, and deploying AI systems that are technically proficient and socially responsible. For example, HCAI practice needs to consider social impacts, ethical considerations, governance, and public engagement.
6) **Interdisciplinary approaches**
The paradigmatic extension of the HCAI approach, including the hierarchical approach and the complementary design metaphors, requires interdisciplinary participation and contributions, including interdisciplinary collaboration and methodology for organizational design, ecosystem design, and social system design [35], [84].

The hHCAI approach, as further illustrated by Figure 4, represents an expansion from a "point" solution (i.e., current HCAI practice focusing on individual human-AI systems) to the "2-D plane" solution (i.e., across multiple human-AI systems within intelligent ecosystems) and then to the "3-D cube" solution that addresses the macro-level sociotechnical systems environment across technical and non-technical subsystems including various organizational and social factors. This expansion mirrors the new requirements placed on STS practices in modern AI, urging solutions beyond the current isolated HCAI approach. Thus, the focus should be on providing comprehensive and systematic human-centered AI solutions for humans and society.

### *C. Implications of the hHCAI framework*

To illustrate the value of the hHCAI approach, Table 4 outlines the need for future HCAI research and practice under the four selected important HCAI topics across the four hierarchical levels [3], [6]. As shown in the table, current HCAI practice takes a siloed approach with a narrow perspective, and truly human-centered AI systems must be implemented across all levels of the HCAI hierarchy.

Table 4 shows that the proposed hHCAI approach, driven by the iSTS framework, demonstrates a comprehensive HCAI approach in a broader sociotechnical context. Thus, HCAI projects are no longer technical projects; they are sociotechnical systems projects requiring interdisciplinary participation and collaboration [35], [50], [79], [84]. The hHCAI approach advocates complementary work across the hierarchical levels from a more systematic perspective. Such a systematic approach will address the weakness in current HCAI practice, enabling us to promote HCAI practice further and deliver end-to-end HCAI solutions.

## VII. Discussion

Both HCAI and STS practices are mutually reinforcing. This paper proposes the iSTS framework to extend traditional STS theory and guide the advances of HCAI practice, maximizing the contributions of the STS approach. The proposed iSTS framework, based on the emerging sociotechnical characteristics demanded by AI and the updated STS design principles, has addressed many issues in alignment with the insights of the previous studies [32], [35], [43], [44], [78], [81], [84], [85], [89], [98], [99], [100]. Furthermore, this paper proposes the hHCAI approach to address the challenges in current HCAI practice driven by the iSTS framework. Both sides share a joint journey to achieve the joint optimization between technical and non-technical subsystems, eventually improving the quality of human work and life.

However, the iSTS framework and the hHCAI approach presented in this paper are the initial working toward extending theory, and they need further research and practice to evolve. Here, we outline recommendations for future work to advance both.

### *A. Enhancing the iSTS framework*

The proposed iSTS framework needs to be enriched for future research and application. Many open questions still need to be addressed; a better understanding of these questions will help enrich the concept and generate practical approaches, eventually building an extended STS framework for the AI era. For example, given AI's learning and evolving features, how can social and AI subsystems co-evolve based on their dynamic interaction and possible collaboration? This includes studying the feedback loops and adaptive behaviors within these systems to understand better how changes in one subsystem can drive changes in the other. How will the interaction and collaboration between humans and AI agents affect human behavior, organizational change, and organizational learning in both organizational and social settings? This research should explore how AI can facilitate or hinder organizational adaptability and resilience [70].

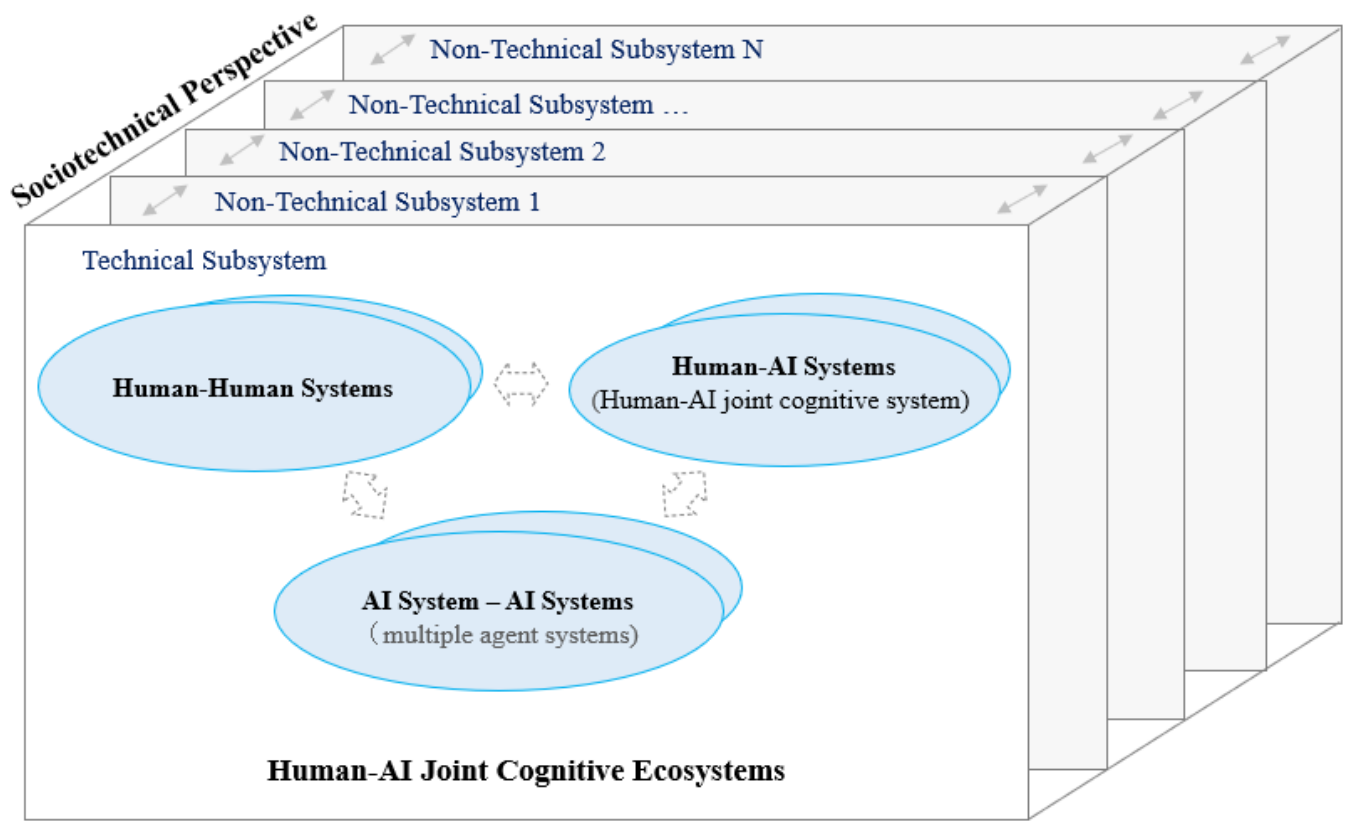

**Fig. 4** Illustration of the relationship across the hierarchical layers of the hHCAI approach (Source: adapted from [44])

Furthermore, we need to determine whether the inclusion of ecosystem and macro-social system levels is solely for the purpose of developing requirements to be considered in technical and organizational design, or if, under certain conditions, these levels should also be directly addressed in the design process. Also, the proposed HCAI approach is presented in a hierarchy fashion. Future work needs to determine the design order of execution, that is, top-down, bottom-up, or combined order, and under certain conditions, the priority of these levels should be addressed by considering the characteristics of the applied domain. Finally, how can we effectively carry out the integrated design and governance of iSTS? Addressing these questions may require developing comprehensive frameworks and ensuring alignment between technical capabilities and social needs, eventually creating a robust STS framework tailored to the complexities of modern AI.

*B. Developing diverse and innovative sociotechnical methods*

While HCAI practice must consider a sociotechnical perspective, STS methods must evolve to keep pace with the times. The iSTS framework emphasizes new design thinking and calls for the development of diverse and innovative sociotechnical methods to support HCAI practice. Over the last 30 years, alternative methods have emerged and are increasingly adopted to drive human-centered solutions from a sociotechnical perspective, such as cognitive work analysis, joint cognitive systems, resilience engineering, organizational ergonomics, and co-design [5], [31], [56], [58], [91], [104], [105], [106], [107], [108], [109], These methods can provide new insights into the complex interplay between humans, organizations, and AI technologies [70].

Practical tools and techniques for the systematic analysis of sociotechnical environments are needed to enable researchers and practitioners to identify potential issues and optimize the integration of AI systems. These tools should facilitate the assessment of human, social, organizational, and technical factors to design human-centered AI systems. Additionally, methods that guide interdisciplinary teams in designing, developing, and deploying AI systems are essential [35], [79], [110]. These methods should support collaborative approaches, combining insights from disciplines such as human factors, computer science, and organizational studies.

*C. Applying iSTS to HCAI practice*

Although iSTS provides valuable insights into HCAI practice, many issues remain regarding its practical application in HCAI research and practice. For instance, how can we effectively apply iSTS in the specific research and development of AI systems? Exploring the application of the iSTS concept in various domains, such as healthcare, transportation, and education, is crucial. This includes conducting case studies to evaluate the effectiveness of iSTS in real-world settings and identifying best practices. Currently, efforts are underway to apply the iSTS framework in various domains. For example, cognitive work analysis was used to analyze the sociotechnical environment of human-automation interaction in aircraft cockpits [111]. The concept of human-AI joint cognitive systems was applied to design human-vehicle co-driving [44]. How can iSTS contribute to ethical AI practice and governance? This involves developing policies and guidelines to ensure AI systems are designed and used in ways that align with societal values and ethical standards from a sociotechnical perspective. Addressing data privacy, security, and public trust issues is essential to ensure that AI systems enhance the quality of life for all members of society. By exploring these questions and conducting comprehensive research, we can better understand how to effectively integrate iSTS into HCAI practice, ensuring that AI empowers humans and aligns with ethical and societal values.

*D. Practicing the hHCAI approach*

We advocate practicing HCAI across multiple levels through a hierarchical approach. This approach requires interdisciplinary collaboration and methodologies, including methods and processes for organizational design, ecosystem design, and social system design [28], [62], [112]. Such collaboration is essential for developing comprehensive solutions that address the challenges of integrating AI into sociotechnical systems.

Creating an organizational and social culture that facilitates interdisciplinary collaboration is crucial. Practicing HCAI across hierarchical levels also helps to promote HCAI further. It is essential to develop metrics and evaluation methods to assess the effectiveness of the hHCAI approach. To comprehensively assess AI integration, these metrics should consider various dimensions, such as user satisfaction, system performance, ethical compliance, and societal impact. As an example, developing an hHCAI maturity mode in an organizational setting will help an organization achieve the ultimate goal of becoming an HCAI-driven organization. The hHCAI maturity model can serve as a diagnosis tool to assess the organization's current state of HCAI practice across various maturity levels, help identify the gaps in the organization to achieve desired goals, and develop an action-based HCAI roadmap for the organization to achieve its ultimate goal.

TABLE 4
NEEDS OF FUTURE HCAI RESEARCH AND PRACTICE FOR SELECTED AREAS ACROSS THE HCAI HIERARCHY BASED ON iSTS
(SOURCE: [3], [6], [17], [28], [32], [45], [49], [61], [74], [96])

| HCAI hierarchy / HCAI areas | Individual human-AI systems | Organizations | Intelligent ecosystems | Social systems |
|---|---|---|---|---|
| Machine behavior management | Human-centered machine learning, user-centered approach (e.g., participatory prototyping and testing) to test and tune algorithms to minimize system bias. | Impacts of machine behavior on organization decision-making, monitoring and evaluation for alignment of organizational goals, the collaborative synergy between AI and workers. | Machine behavior evolution across multiple AI systems within an ecosystem, human-AI trust ecosystem, conflict resolution mechanisms across vendors, decision-making authority across systems. | Effects of the macro-social environment on machine behavior, machine behavior evolvement in social interaction, ethics of machine behavior across cultures. |
| Human-led collaboration with AI | Human-AI function reallocation, collaboration-based cognitive UI design and user validation, mutual trust, decision-making sharing, human controllability. | Work system design (e.g., role, process), automation / routine task design, skill development, organization structure change, job enrichment, collaboration for HCAI solutions. | Human-AI collaboration across AI systems, adaptation, co-learning, and evolution mechanisms across multiple AI systems and organizations within an ecosystem. | The social interaction mechanism of human-AI teams in the social environment, the impact of social responsibility on human-AI collaboration. |
| Explainable AI (XAI) | Human-centered explainable AI, explainable UI modeling and design, UI visualization, application of psychological explainability theories. | Impacts of AI explainability on organization decision-making, transparency and trust, compliance and accountability of AI-driven decisions, risk management. | XAI across multiple AI systems, decision-making synchronization, XAI-based interoperability across systems, integrity and trust across systems, evolvement and adaptation of XAI systems. | The relationship between explainable AI and public trust, acceptance, culture, and ethics, human-centered explainable AI from a sociotechnical perspective. |
| Ethical AI | System output fairness, data privacy protection, accountability tracking, algorithm governance, reusable ethical AI code. | Ethical AI guidelines and governance for organizations, organizational culture, developer skill training. | Ecosystem-wide fairness and bias mitigation, cross-organizational data ethics and privacy, ethical impact assessment and auditing, human-AI trust in ecosystems. | Ethical AI implications in society, social impacts, social trust in AI, AI policy and regulation, impact on the labor market and workforce skills, policy and regulations. |

Future research and practice can further advance the iSTS framework and the hHCAI approach by addressing these recommendations. This will ensure that AI systems are designed, developed, and deployed in ways that truly benefit humans and society. Ongoing efforts in this direction will help create a more human-centered AI ecosystem, fostering positive outcomes for individuals, organizations, and society.

## VIII. CONCLUSION

In this paper, we have introduced the iSTS framework to extend traditional STS theory with respect to modern AI, through a literature review, comparative analysis, identification of emerging sociotechnical characteristics and needs, and development of updated STS design principles. iSTS aims to address the complex challenges posed by AI technology in the sociotechnical environment. Our approach emphasizes human-centered joint optimization between social and AI subsystems across four hierarchical levels: individual human-AI systems, organizations, ecosystems, and social systems.

Based on iSTS, we have further proposed the hHCAI approach. The hHCAI approach provides a structured approach across the levels of individual human-AI systems, organizations, ecosystems, and social systems to integrate AI design, development, and deployment into sociotechnical systems, ensuring that AI technologies align with human values, needs, and societal goals. The framework also helps address the challenges in current HCAI practice.

Collaboration between the HCAI and STS communities is essential for developing truly human-centered AI systems that maximize AI's benefits while minimizing its negative impacts. The proposed iSTS framework and the hHCAI approach represent a paradigmatic extension of current STS and HCAI practices, offering actionable insights and recommendations for future research and practical application.

Future work should further enhance the iSTS framework and the hHCAI approach, develop innovative sociotechnical methods, and apply iSTS and hHCAI to practice. By fostering interdisciplinary collaboration and systematically addressing the sociotechnical environment, we can advance HCAI practice and deliver AI systems that better serve humans and society. The journey towards achieving the human-centered joint optimization between humans, organizations, society, and AI is ongoing, and this paper aims to contribute to that critical endeavor.

## ACKNOWLEDGMENT

The authors appreciate the insightful comments from three anonymous reviewers and the journal’s editor-in-chief, Professor Katina Michael, on an earlier draft of this article. These insights have helped improve the quality of this article.

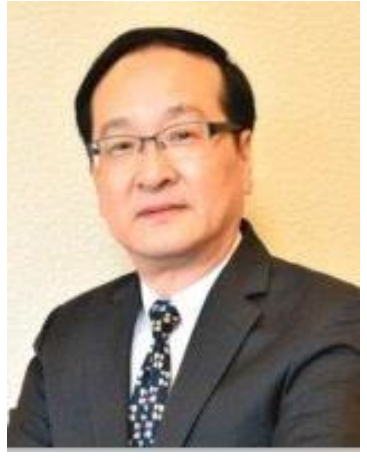

**Wei Xu** is a Professor of Human Factors and Psychology in the Department of Psychology and Behavioral Sciences and the Center for Psychological Sciences at Zhejiang University, China. He is an elected fellow of the International Ergonomics Association, the Human Factors and Ergonomics Society, and the Association for Psychological Science. He received his Ph.D. in Human Factors and M.S. in Computer Science from Miami University in 1997. He serves as an Associate Editor for IEEE Transactions on Human-Machine Systems. His research interests include HCAI, human-AI interaction, and aviation human factors.

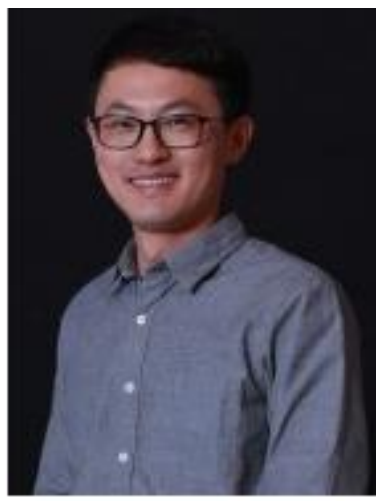

**Zaifeng Gao** is a Professor of Psychology and Human Factors in the Department of Psychology and Behavioral Sciences at Zhejiang University, China. He received his Ph.D. in Psychology from Zhejiang University, China, in 2009. His research interests include engineering psychology, autonomous driving, and cognitive psychology.